\documentclass[pre,aps,preprint,showpacs]{revtex4}
\usepackage{revsymb,latexsym,amssymb,amsmath,comment}
\usepackage{graphicx,psfrag,subfigure,stmaryrd}

\begin{document}
\author{J. M. Wesselinowa }
\email{julia@phys.uni-sofia.bg}
\affiliation{University of Sofia, Department of Physics\\ Blvd. J. Bouchier 5, 
1164 Sofia, Bulgaria}
\author{T. Michael, S. Trimper,}
\email{trimper@physik.uni-halle.de}
\author{K. Zabrocki}
\affiliation{Fachbereich Physik,
Martin-Luther-Universit\"at, D-06099 Halle Germany}
\title{Influence of layer defects on the damping in ferroelectric thin films}
\date{\today }
\begin{abstract}
A Green's function technique for a modified Ising model in a transverse field is applied,
which allows to calculate the damping of the elementary excitations and the phase transition 
temperature of ferroelectric thin films with structural defects. Based on an analytical 
expression for the damping function, we analyze its dependence on temperature, film thickness 
and interaction strength numerically. The results demonstrate that defect layers in ferroelectric 
thin films, layers with impurities or vacancies as well as layers with dislocations are able to induce 
a strong increase of the damping due to different exchange interactions within 
the defect layers. The results are in good agreement with experimental data for thin ferroelectric 
films with different thickness. 
\end{abstract}

\maketitle

\section{Introduction}
Defects in crystals influence the physical properties of almost all materials significantly. In
the last decades, defect engineering has been developed as a part of modern semiconductor
technology. There is also an increasing interest in studying defects and their related
strain fields in ferroelectrics, in particular in low dimensional systems \cite{1}.
The influence of defects on the phase transition is one of key-like topics in the context of 
recent studies of ferroelectrics. The permanent attention attracted to ferroelectric (FE) material, is related 
to its role played in real ferroelectrics concerning fundamental problems as well as the broad 
variety of applications based on thin FE films, especially due to their suitability for application 
in nonvolatile FE random access memories \cite{1a}. It is well known that localized spin excitations can arise
in FE and magnetic systems with broken spatial translational symmetry, i.e. due to the presence of boundary
surfaces, interfaces, impurities and other defects in the crystal. These localized modes
appearing in the above mentioned structures are experimentally observed using the neutron
and Raman scattering \cite{2,3} or, more recently, far infrared absorption \cite{4}. From more theoretical 
point of view the Green's function theory has been extensively employed to describe these modes 
in semi-infinite transverse Ising models \cite{5} and FE thin films \cite{6}. In addition to bulk spin waves 
there may occur localized modes associated with the impurity layer \cite{5} and with the surface \cite{6}. 
More phenomenologically the influence of molecular impurity ions on FE phase transitions is studied by 
Vikhin and Maksimova using the Landau theory \cite{7}.

The relaxation (damping) of polarization motion is a result of microscopic fluctuations of the 
spins by means a coupling to other degrees of freedom such as phonons, impurities, defects etc. The 
mechanism behind can be originated by different reasons. Once there could be realized 
a direct coupling of the elementary spin wave excitations to a thermal bath or to phonons manifested 
by a spin-phonon scattering. Otherwise more complicated microscopic processes, such as "slow relaxing" 
impurities could control the fluctuations and the relaxation processes of the system. In that case the 
impurity modes are damped which leads to a relaxation of the polarization motion due to a direct coupling. 
In the present paper we want to study such a mechanism in detail based on an Ising model in a transverse field. 
There are two kind of relaxation processes: inherent (pure bulk material) and non-inherent
(due to defects, impurities, surface, finite size). Each microscopic relaxation mechanism predicts its own
temperature and frequency dependencies. The damping can also depend on defect/impurity concentration
and the sample size. Using X-ray diffraction it is shown that too many structural defects can produce an 
excessive line broadening \cite{8}, e.g. defects contribute to the line width and consequently the damping 
increases. Doping of thin films with different ions yields to variable increase of the damping \cite{9,10,11}. 
Ingvarsson et al. \cite{12} showed that the relaxation mechanism in thin films could be realized in analogy to 
the bulk relaxation, where the phonon scattering process in the bulk should be replaced by surface and defect 
scattering in thin films.

A first attempt to find out the influence of defects on the polarization of FE thin films
are elucidated by Alpay et al. \cite{13,14}. Based on the Ising model in a transverse field and
using the Green's function theory Wesselinowa et al. \cite{15} had been successfully in calculating 
the spin-wave energies, the polarization and the phase transition temperature for a FE thin film 
with different structural defect layers. The aim of the present paper is to study the damping 
effects and the relaxation times in FE thin films with different structural defect layers. To that purpose 
a previous paper \cite{16} is extended and modified in order to obtain the temperature dependence of the 
soft modes and their damping within FE thin films. It was obtained that with increasing film thickness the
frequency increases likewise and the damping is larger for thin films than that for the bulk material. This is
in agreement with different experimental observations \cite{17,18,19}. Using Raman spectroscopy the temperature 
dependence of the phonon modes for thin FE films of PbTiO$_3$ is discussed by Taguchi et al. \cite{17} and 
Fu et al. \cite{18}. It was shown that in comparison with the single crystal spectra, the Raman frequencies 
for the thin film are shifted remarkably to low frequencies and that the Raman lines are broadened. Similar results 
were obtained for PbTiO$_3$ fine particles by Ishikawa et al. \cite{19}.

\section{The Model and the Matrix Green's Function}
Let us consider a three dimensional ferroelectric system on a simple cubic lattice composed
of $N$ layers in z-direction. The layers are numbered by $n=1,...N$, where the layers $n=1$ and
$n=N$ represent the two surfaces of the system. The bulk is established by the remaining
$(N-2)$ layers. The specific surface effects are included by additional coupling parameters
between bulk and surface layers. In particular, we start with the Hamiltonian of the Ising
model in a transverse field which includes both, bulk and surface properties:
\begin{equation}
H=-\frac{1}{2}\sum_{ij}J_{ij}S^z_iS^z_j-\Omega_b\sum_{i\epsilon b}S^x_i-
\Omega_s\sum_{i\epsilon s}S^x_i\,,
\label{ham}
\end{equation}
where $S^x$ and $S^z$ are components of spin-$\frac{1}{2}$ operators, $\Omega_b$ and $\Omega_s$ 
represent transverse fields in the bulk and surface
layers, and the sums are over the internal and surface lattice points,
respectively. $J_{ij}$ is an exchange interaction between spins at
nearest-neighbor sites $i$ and $j$, and $J_{ij}=J_s$ between spins on the
surface layer, otherwise it is $J_b$.
We assume that one or more of the layers can be defect, since
$J_d$ and $\Omega_d$ denote the exchange
interaction and the transverse field of the defect layer. The ordered phase is 
characterized by the non zero mean values $\langle S^x \rangle \neq 0$ and $\langle S^z \rangle \neq 0$. 
Hence it is appropriate to choose a new coordinate system by rotating the original one, used in 
Eq.~(\ref{ham}), by the angle $\theta$ in the $xz$ plane \cite{20}. The rotation angle $\theta$ 
is determined by the requirement $\langle S^{x'} \rangle = 0$ in the new coordinate system.

The retarded Green's function to be calculated is defined as
\begin{equation}
G_{ij}(t)=\ll{S^+_i(t);S^-_j(0)}\gg,
\end{equation}
where $S^+$ and $S^-$ are the spin-$\frac{1}{2}$ operators in the rotated system.
On introducing the two-dimensional Fourier transform $G_{n_in_j}({\bf k}_
\parallel,\omega)$, one has the following form:
\begin{equation}
\ll{S^+_i;S^-_j}\gg_{\omega}=\frac{\sigma}{N'}\sum_{{\bf k}
_{\parallel}}\exp(i{\bf k}
_{\parallel}({\bf r}_i-{\bf r}_j))G_{n_in_j}({\bf k}_{\parallel}, \omega),
\label{gf}
\end{equation}
where $N'$ is the number of sites in any of the lattice planes, ${\bf r}_i$
and $n_i$ represent the position vectors of site $i$ and the layer index,
respectively, ${\bf k}_{\parallel}=(k_x,k_y)$ is a two-dimensional wave vector
parallel to the surface. The summation is taken over the Brillouin zone.

As a result the equation of motion for the Green's function Eq.~(\ref{gf}) of the ferroelectric
thin film for $T \leq T_c$ has the following matrix form:
\begin{equation}
{\bf H}(\omega){\bf G}({\bf k}_{\parallel},\omega)={\bf R},
\label{ham1}
\end{equation}
where ${\bf H}$ can be expressed as:
\begin{displaymath}
\mathbf{H} =
\left( \begin{array}{ccccccc}
\omega-V_1+i\gamma_1 &   k_1      &    0       &    0   &    0   &    0   & \ldots \\
    k_2    & \omega-V_2+i\gamma_2 &   k_2      &    0   &    0   &    0   & \ldots \\
   0     &   k_3      & \omega-V_3+i\gamma_3 &   k_3  &    0   &    0   & \ldots \\
      \vdots  & \vdots & \vdots & \vdots & \vdots & \vdots & \ddots\\
0 & 0 & 0 & 0 & 0 & k_N & \omega-V_N+i\gamma_N \\
\end{array} \right)
\end{displaymath}
with
\begin{eqnarray*}
k_n&=& J_b \sigma_n \sin^2{\theta_n}, \qquad n=1,...,N, \\
V_1&=&2\Omega_s \sin{\theta}_1 + \frac{1}{2} \sigma_1 J_s \cos^2{\theta}_1-
\frac{\sigma_1 J_s}{4} \sin^2{\theta}_1 \gamma({\bf k_{\parallel}})+
J_b\sigma_2 \cos^2{\theta}_2, \\
V_2&=&2\Omega_b \sin{\theta}_2 + \frac{1}{2} \sigma_2 J_b \cos^2{\theta}_2-
\frac{\sigma_2 J_b}{4} \sin^2{\theta}_2 \gamma({\bf k_{\parallel}})+J_s \sigma_1
\cos^2{\theta}_1 + J_b\sigma_3 \cos^2{\theta}_3, \\
V_n&=&2\Omega_n \sin{\theta}_n + \frac{1}{2} \sigma_n J_b \cos^2{\theta}_n-\frac{\sigma_n J_n}
{4} \sin^2{\theta}_n \gamma({\bf k_{\parallel}})+J_{n-1}\sigma_{n-1} \cos^2{\theta}_{n-1}
\\ &+&J_{n+1}\sigma_{n+1} \cos^2{\theta}_{n+1}, \\
V_N&=&2\Omega_s \sin{\theta}_N + \frac{1}{2} \sigma_N J_s \cos^2{\theta}_N-
\frac{\sigma_N J_s}{4} \sin^2{\theta}_N \gamma({\bf k_{\parallel}})+J_b\sigma_{N-1}
\cos^2{\theta}_{N-1},\\
\gamma_n&=&\frac{\pi}{2N^2}\sum_{{\bf p}_{\parallel},{\bf q}_{\parallel}}
\Big[(\bar{V}_n({\bf q}_{\parallel},{\bf k}_{\parallel}-{\bf q}_{\parallel})+
\bar{V}_n({\bf k}_{\parallel}-{\bf p}_{\parallel}-{\bf q}_{\parallel},
{\bf p}_{\parallel}+{\bf q}_{\parallel}))^2\nonumber\\
&*&[\bar{n}_n({\bf p}_{\parallel})
(\sigma_n+\bar{n}_n({\bf p}_{\parallel}+{\bf q}_{\parallel})+
\bar{n}_n({\bf k}_{\parallel}-{\bf q}_{\parallel}))\nonumber\\
&-&\bar{n}_n({\bf p}_{\parallel}+{\bf q}_{\parallel})\bar{n}_n
({\bf k}_{\parallel}-{\bf q}_{\parallel})]\nonumber\\
&*&\delta(\epsilon_n({\bf k}_{\parallel}-{\bf q}_{\parallel})+
\epsilon_n({\bf p}_{\parallel}+
{\bf q}_{\parallel})
-\epsilon_n({\bf p}_{\parallel})-\epsilon_n({\bf k}_{\parallel}))\nonumber\\
&+&[(J_{n-1}\gamma({\bf q}_{\parallel})\cos^2{\theta}_{n-1})^2+
(J_{n-1}\gamma({\bf k}_{\parallel}-{\bf p}_{\parallel}-{\bf q}_{\parallel})
\cos^2{\theta}_{n-1})^2]\nonumber\\
&*&[\bar{n}_{n-1}({\bf p}_{\parallel})
(\sigma_{n-1}+\bar{n}_{n-1}({\bf p}_{\parallel}+{\bf q}_{\parallel})+
\bar{n}_{n-1}({\bf k}_{\parallel}-{\bf q}_{\parallel}))\nonumber\\
&-&\bar{n}_{n-1}({\bf p}_{\parallel}+{\bf q}_{\parallel})\bar{n}_{n-1}
({\bf k}_{\parallel}-{\bf q}_{\parallel})]\nonumber\\
&*&\delta(\epsilon_{n-1}({\bf k}_{\parallel}-{\bf q}_{\parallel})+
\epsilon_{n-1}({\bf p}_{\parallel}+
{\bf q}_{\parallel})
-\epsilon_{n-1}({\bf p}_{\parallel})-\epsilon_{n-1}({\bf k}_{\parallel}))\nonumber\\
&+&[(J_{n+1}\gamma({\bf q}_{\parallel})\cos^2{\theta}_{n+1})^2+
(J_{n+1}\gamma({\bf k}_{\parallel}-{\bf p}_{\parallel}-{\bf q}_{\parallel})
\cos^2{\theta}_{n+1})^2]\nonumber\\
&*&[\bar{n}_{n+1}({\bf p}_{\parallel})
(\sigma_{n+1}+\bar{n}_{n+1}({\bf p}_{\parallel}+{\bf q}_{\parallel})+
\bar{n}_{n+1}({\bf k}_{\parallel}-{\bf q}_{\parallel}))\nonumber\\
&-&\bar{n}_{n+1}({\bf p}_{\parallel}+{\bf q}_{\parallel})\bar{n}_{n+1}
({\bf k}_{\parallel}-{\bf q}_{\parallel})]\nonumber\\
&*&\delta(\epsilon_{n+1}({\bf k}_{\parallel}-{\bf q}_{\parallel})+
\epsilon_{n+1}({\bf p}_{\parallel}+
{\bf q}_{\parallel})
-\epsilon_{n+1}({\bf p}_{\parallel})-\epsilon_{n+1}({\bf k}_{\parallel}))\Big],\nonumber\\
\bar{V}({\bf {q_{\parallel},k_{\parallel}-q_{\parallel}}})&=&J({\bf q_{\parallel}})\cos^2{\theta}-\frac{1}{2}
J({\bf {k_{\parallel}-q_{\parallel}}})\sin^2{\theta},\nonumber\\
\gamma({\bf k}_{\parallel})&=&\frac{1}{2}(\cos(k_x a)+\cos(k_y a)).
\end{eqnarray*}
Here we have introduced the notations $J_1 \equiv J_N = J_s$, $J_n=J_b$ for $n=2,3,4,...,N-1$,
$\Omega_1=\Omega_N=\Omega_s$, $\Omega_n=\Omega_b\, n=2,3,4,...,N-1),\, J_0=J_{N+1}=0$. The quantity $\sigma(T)$
is the relative polarization in the direction of the mean field and is equal to
$2\langle S^{z'} \rangle$. For the rotation angle $\theta$ we have the following
two solutions in the generalized Hartree-Fock approximation:
\begin{eqnarray*}
&1.& \cos{\theta}=0, \quad i.e.\quad  \theta=\frac{\pi}{2}, \quad\quad \mbox{if} \quad
T \geq T_c\,;\nonumber\\
&2.& \sin{\theta}=\frac{4\Omega}{\sigma J}=\frac{\sigma_c}{\sigma},
\quad\quad\quad\quad\quad \mbox{if} \quad T\leq T_c\,.
\end{eqnarray*}

In order to obtain the solutions of the matrix Eq.~(\ref{ham1}), we introduce the
two-dimensional column matrices, ${\bf G}_m$ and ${\bf R}_m$, where the elements
are given by $({\bf G}_{n})_m = G_{mn}$ and
$({\bf R}_{n})_m = \sigma _n \delta _{mn}$, so that Eq.~(\ref{ham1}) yields
\begin{equation}
{\bf H}(\omega ) {\bf G}_n ={\bf R}_n.
\label{ham2}
\end{equation}
From Eq.~(\ref{ham2}), $G_{nn}(\omega )$ is obtained as:
\begin{equation}
G_{nn}(\omega ) = \frac{\lvert H_{nn}(\omega ) \rvert}{\mid H(\omega )\mid}.
\end{equation}
The quantity $ \lvert H_{nn}(\omega ) \rvert $ is the determinant made by replacing the $n$-th
column of the determinant $\lvert H(\omega ) \rvert $ by $R_n$. The poles $\omega _n$ of the
Green`s function $G_{nn} (\omega )$ can be calculated by solving
$\lvert H(\omega ) \rvert = 0$.

The relative polarization of the $n$-th layer is given by
\begin{equation}
\sigma_n=\Big(\frac{\sigma_nJ_n}{2N}\sum_{{\bf k}_{\parallel}}\frac{1-
0.5\sin^2{\theta_n} \gamma({\bf k}_{\parallel})}{\omega_n}
\coth{\frac{\omega_n}{2T}}\Big)^{-1}.
\label{eq7}
\end{equation}
Eq.~(\ref{eq7}) has to be numerically calculated. Due to the assumption of
symmetrical surfaces, there are $\frac{1}{2} N$ layer polarizations, which have
to be solved self-consistently. In order to obtain the dependence of the Curie
temperature $T_c$ on the film thickness $N$, we let all $\sigma$'s be zero in
Eq.~(\ref{eq7}) and solve the expressions self-consistently.

\section{Numerical Results and Discussion}
\begin{figure} [!ht]
\centering
\psfrag{si}[][][1.3]{${\sigma}$}
\psfrag{Temp}[][][0.75]{$T [K]$}
\includegraphics[scale=1.0]{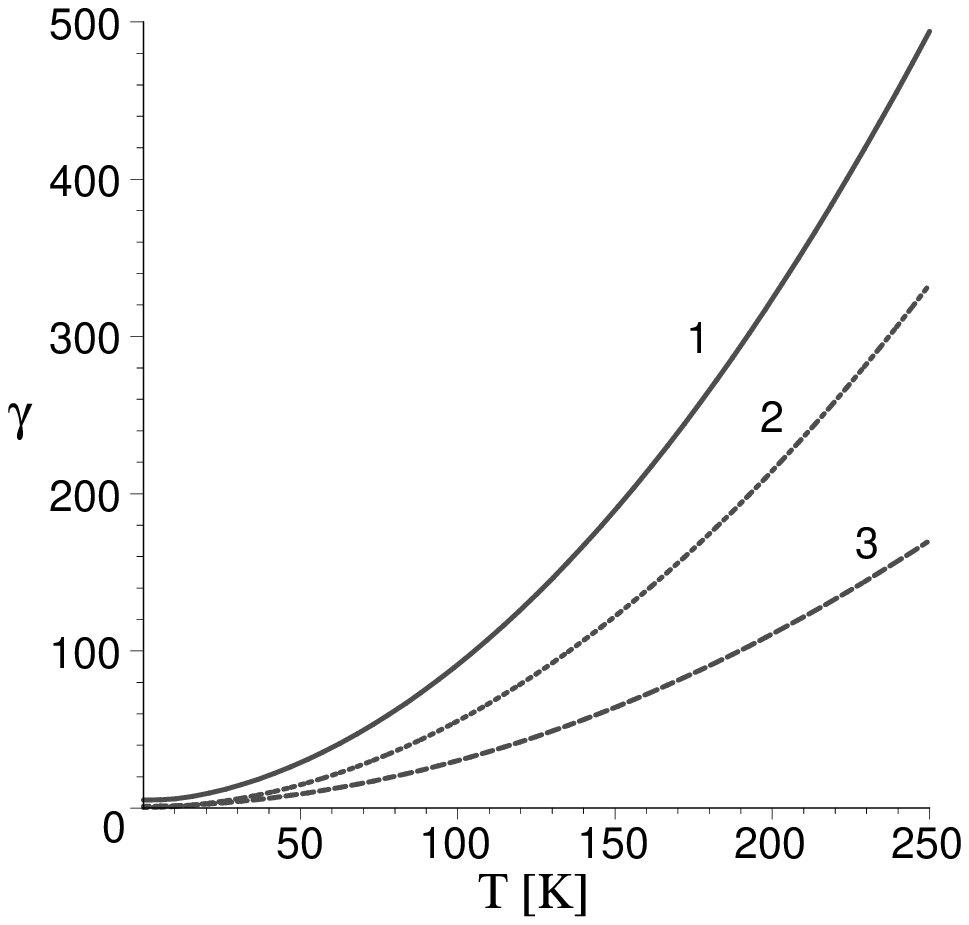}
\caption{Temperature dependence of the damping $\gamma$ in cm$^{-1}$ for a FE thin film with $J_b = 495$\,K,
$\Omega_b = 20$\,K, $J_s = 900$\,K, $\Omega_s=\Omega_b$, $N = 7$ and different $J_d$-values:
(1): $J_d = 300$\,K;\,(2): $1000$\,K;\, (3): $495$\,K.}
\label{Fig.1}
\end{figure}

In this section we shall present the numerical calculations of our
theoretical results taking the following model parameters: $J_b =
495$\,K, $\Omega_b= 20 $\,K. We have calculated the temperature dependence of 
the damping from $\gamma=\frac{1}{N}\sum_n{\gamma_n}$ for a simple cubic thin film 
($\bf k_{\parallel} = 0)$ and for different values of the exchange interaction 
constants. The numerical results expose some interesting and novel
characteristics in the damping values in comparison to the case of
FE thin films without defects. The results for film thickness
$N=7$ and different exchange interaction parameters in the defect
layer $J_d$ are presented in Fig.~\ref{Fig.1}. We consider firstly the case
where the middle layer is defect, which is possibly the case when
the layer has vacancies or impurities with smaller radius and
larger distances between them. The exchange interaction $J_d$ is
smaller than the value of the bulk interaction $J_b=495$K and has
the value $J_4=J_d=300$\,K (Fig.~\ref{Fig.1}, curve 1). The damping is larger
than in the case without defects, $J_4 = J_b$ (Fig.~\ref{Fig.1}, curve 3). This
is in agreement with the experimental data \cite{9,10,11,12}.  Unfortunately,
there are more experimental data for damping of magnetic thin
films and not for ferroelectric thin films. The damping increases
with increasing temperature, $T \rightarrow T_c$ of the thin film.
This could explain the observed experimental results of Raman
scattering from ferroelectric thin films that the line shapes of
the film become broad as the temperatures approaches $T_c$ \cite{17,18,19}. 
The critical temperature decreases due to the smaller $J_d$ value. 
From Eq.~(\ref{eq7}) by solving $\sigma (T_c) = 0$ we get the 
Curie temperatures: $T_c(\mbox{bulk}) = 219.5$\,K and $T_c(J_d = 300)$\,K $ = 213.8$\,K. 
Pontes et al. \cite{20} carried out dielectric and Raman spectroscopy studies and obtained 
that with addition of Sr to PbTiO$_3$, the phase transition temperature decreases with the
increasing Sr concentration.

For the case where $J_4 = J_d = 1000$\,K (Fig.~\ref{Fig.1}, curve 2), i.e. $J_d$ is larger than the
value of the bulk interaction constant $J_b$, (for example when the impurities have a
larger radius compared with the constituent ions) the damping is again larger in comparison to the case 
without defects, $J_d=J_b$. Experimentally one observes an increase of the damping in presence of defects 
\cite{9,10,11,12}. The value of $T_c$ of the film is enhanced in comparison to the bulk value without defects 
due to larger $J_d$ values. It results: $T_c(J_d = 1000)$\,K $ = 228.5$\,K. This is the opposite behavior 
compared to the case of $J_d = 300$\,K, $J_d < J_b$. The second case, where $J_d > J_b$ could explain the 
experimentally obtained increase in the phase transition temperature by the substitution of impurities, 
such as Nd, V, W and Nb in the layers of BTO films \cite{21,22,23} or by increasing Ca contents in SBT 
thin films \cite{24}.

We obtain that the critical temperatures of the FE phase transition are increased or decreased due to different 
exchange interactions in the defect layers. Our results are in qualitatively agreement with
the experimental data of Noguchi et al.\,\cite{25}. They have studied SBT films
($T_c=295^0$\,C) by the substitution of rare earth cations of La, Ce, Pr, Nd and Sm as well as
Bi at the A site (Sr site) with Sr vacancies and have shown, that La modification induces
soft behavior (lower $E_c$ and lower $P_r$), while a large amount of Nd and Sm substitution
results in a very high $E_c$ (hard), as a result of defect engineering of both Sr and oxide
vacancies. For not only SBT, but also other bismuth layer-structured ferroelectric (BLSF) $T_c$ is
strongly influenced by $r_i$ of A-site cations, and BLSF with smaller A-site cations
(Ca$^{2+}$) tend to show a higher $T_c$ (420$^0$\,C). The same amount of larger Ba$^{2+}$
brings about a relaxation of FE distortions and leads to a decrease of $T_c$ to 120$^0$C.
The substitution of La led to a marked decrease in $T_c$ to 180$^0$\,C ($x$=0.5),
because the induced A-site vacancies weaken the coupling between neighboring
BO$_6$ octahedra \cite{25,26}. This result corresponds in our calculations to the case of
smaller values of the interaction constant in the defect layer $J_d<J_b$. For
La-modified PbTiO$_3$, $T_c$ decreased significantly, too, with an increase in La content
\cite{27}. For Bi-SBT $T_c$ rose strongly to 405$^0$\,C ($x$=0.2) \cite{24}.
The increase in $T_c$ by Bi substitution is an opposite tendency to that in the case for La-SBT. The dominant role
plays here the bonding characteristics with oxide ions. The influence of the orbital
hybridization on $T_c$ is very large, and Bi substitution resulted in a higher $T_c$ \cite{25}.
This experimental result can be described qualitatively good using in our model exchange
interaction parameters for the defect layer $J_d > J_b$.

\begin{figure} [!ht]
\centering
\psfrag{si}[][][1.3]{${\sigma}$}
\psfrag{Temp}[][][0.75]{$T [K]$}
\includegraphics[scale=1.0]{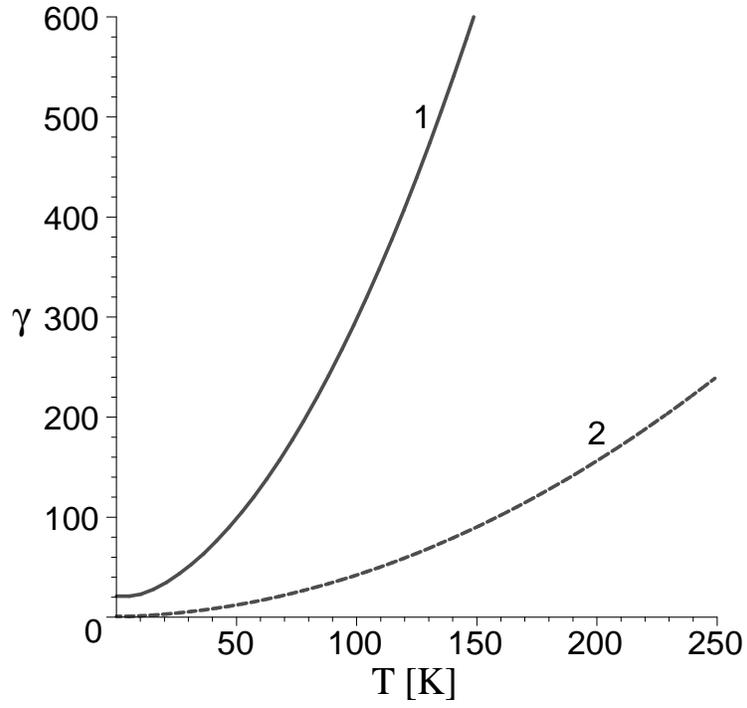}
\caption{Temperature dependence of the damping $\gamma$ in cm$^{-1}$ for $J_b = 495$\,K, $\Omega_b = 20$\,K,
$J_s = 300$\,K, $\Omega_s = \Omega_b$, $N = 7$ and different defect layers: (1): $J_3 = J_4 = J_5 = J_d = 600$\,K; 
(2): $J_4 = J_d = 600$\,K}
\label{Fig.3}
\end{figure}

The damping depends also on the number of inner defect layers. This is shown
in Fig.~\ref{Fig.3} for a thin film with $N=7$ layers. It is demonstrated that
for $J_s = 300 K$ with an increasing number of defect layers (curve 1 - $J_3=J_4=J_5=J_d = 600K$, curve 2 -
$J_4=J_d= 600 K$) the damping increases, i.e. we obtain a line broadening. This is in agreement with the
experimental data of Boulle et al. \cite{8} who argued that too many structural defects can produce excessive line broadening.

\begin{figure} [!ht]
\centering
\psfrag{si}[][][1.3]{${\sigma}$}
\psfrag{Temp}[][][0.75]{$T [K]$}
\includegraphics[scale=1.0]{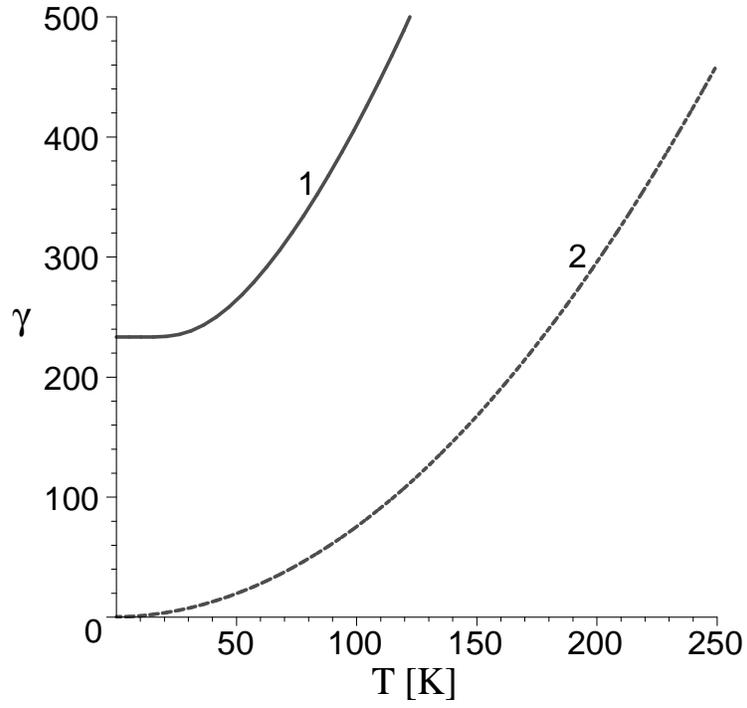}
\caption{Temperature dependence of the damping $\gamma$ in cm$^{-1}$ for $J_b = 495$\,K,\,$\Omega_b = 20$\,K,\,
$J_s = 600$\,K,\,$J_d = 1000$\,K,\,$\Omega_s = \Omega_b$ and different film thickness: (1): $N = 7$;\, (2): $N = 9$.}
\label{Fig.4}
\end{figure}

In Fig.~\ref{Fig.4} we have studied a ferroelectric thin film with different film thickness where one layer, the
middle one is defect, $J_d=1000 K$ ($N=7$ - curve 1) and ($N = 9$ - curve 2).
The damping increases with decreasing of the film thickness, what is in agreement with our previous calculations
without defects \cite{16} and with the experimental data \cite{17,18,19}. 
The thinner the films the larger is the damping.

\section{Conclusions}
Based on a modified transverse Ising model and using a Green's function technique,
the damping and the phase transition temperature for ferroelectric
thin films with structural defects are calculated for the first time. The dependence on temperature, film thickness
and interaction constants is discussed. It is shown that defect layers in FE thin films, layers
with impurities or vacancies, as well as layers with dislocations, can induce strong increase or decrease of 
the critical temperature of FE phase transition due to different exchange interactions in the
defect layers. We obtain that the damping in the thin films is always larger in comparison to the 
corresponding bulk values or to films without defects. There are different mechanisms
in thin films which contribute additive to the damping, due to spin-spin, spin-phonon, surface, defects scattering etc.
The damping in thin films is greater compared to the bulk case for example due to surface effects \cite{16} then due to
spin-phonon interactions \cite{28} and to defects which are considered in the present paper. The damping is related
to the width of the half-maximum for the Raman scattering lines. It can be concluded,
that if we want to obtain and to explain
the large line broadening effects observed experimentally in thin films and nanoparticles \cite{17,18,19} all contributions
to the damping must be taken into account and that the defects play an important role.

\begin{acknowledgments} 
One of us (J. M. W.) is grateful to the Deutsche Forschungsgemeinschaft 
for financial support. This work is supported by the SFB 418 and by the cluster of excellence. 
\end{acknowledgments}

\newpage
\clearpage

\clearpage
\clearpage

\end{document}